\documentclass[showpacs,amsmath,twocolumn,floatfix]{revtex4-1}

\usepackage{subfigure}
\usepackage{graphicx}
\usepackage{amssymb}
\usepackage{amsmath}
\usepackage{bbm}
\usepackage[backref=none]{hyperref} 
\usepackage[usenames,dvipsnames]{color}
\usepackage{stmaryrd}
\usepackage{rotating}
\DeclareMathAlphabet{\cali}{OMS}{zplm}{m}{n}

\definecolor{MyDarkBlue}{rgb}{0,0.1,0.7}
\hypersetup{pdfborder={0 0 0},colorlinks,breaklinks=true,
  urlcolor={MyDarkBlue},citecolor={MyDarkBlue},linkcolor={MyDarkBlue}
}
\newcommand\mydoi[1]{\href{https://doi.org/#1}{doi:#1}}

\newcommand\ii{\mathrm{i \,}}
\newcommand\dd{\delta\:\!}

\begin{document}

\title{Particle-hole symmetry of charge excitation spectra in the paramagnetic
  phase of the Hubbard model} 
\author{Vu Hung Dao}
\affiliation{Normandie Univ, ENSICAEN, UNICAEN, CNRS, CRISMAT, 14000 Caen, France}
\author{Raymond Fr\'esard} 
\email[Corresponding author: ]{Raymond.Fresard@ensicaen.fr}
\affiliation{Normandie Univ, ENSICAEN, UNICAEN, CNRS, CRISMAT, 14000 Caen,
  France}


\begin{abstract}
The Kotliar and Ruckenstein slave-boson representation of the Hubbard model
allows to obtain an approximation of the charge dynamical response function
resulting from the Gaussian fluctuations around the paramagnetic saddle-point
in analytical form. Numerical evaluation in the thermodynamical limit yields
charge excitation spectra consisting of a continuum, a gapless collective mode
with anisotropic zero-sound velocity, and a correlation induced high-frequency
mode at $\omega\approx U$. In this work we show that this analytical expression
obeys the particle-hole symmetry of the model on any bipartite lattice with
one atom in the unit cell. Other formal aspects of the approach are also
addressed. 
\end{abstract}

\maketitle
\section{Introduction}
\label{sec:int}
Most peculiar
properties of transition metal oxides that attract a lot of attention
are believed to result from strong electronic correlations. A great variety of
physical phenomena has been evidenced \cite{Ima98}, with prominent examples
being the 
striking metal-to-insulator transitions in vanadium sesquioxide
\cite{McW73,Hel01,Lim03,Gry07}, high-$ {\rm T_c} $ superconductivity in the
cuprates \cite{Bed86,Sch93}, non-Fermi liquid behavior in the vanadates
\cite{Ina95}, stripes in nickelates~\cite{Sac95n,Tra17} and 
cuprates~\cite{Tra95c,Tra17b}, or the colossal magnetoresistance observed in the
manganites \cite{Hel93,Tom95,Rav95,Mai95}. In addition, a whole series of
promising materials for thermoelectric applications has been discovered
\cite{Ter97,Mas00,Mat01,Mic07,Ohta07,Mai09b,Wang13}.

\begin{figure}[h]
  \includegraphics[clip=true, width=0.48\textwidth]{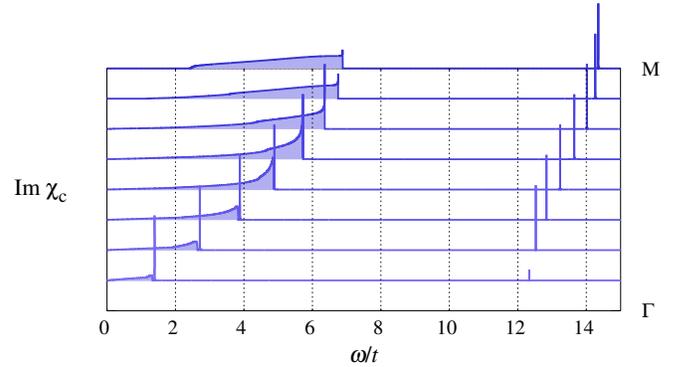}
  \caption{(Color online) Spectrum of the inelastic charge response function
    ${\rm Im} \chi_c$ at low temperature $T$ for different momenta along the
    path linking $\Gamma=(0,0)$ to ${\rm M}=(\pi,\pi)$, at coupling $U=10t$
    and doping from half-filling $\delta=\pm 0.5$. Note the peak of the
    zero-sound collective mode above the upper edge of the continuum, and the
    UHB mode peak at higher energy.  
  Parameter: $T=t/100$.}
  \label{fig:fs}
\end{figure}

The infancy of the microscopical modeling of strongly correlated systems
dates to the early sixties with the introduction of the so-called one-band
Hubbard Model \cite{Hub63,Hub64,Kan63,Gut63}, 
\begin{equation}
H = -t \sum_{\langle i,j\rangle ,\sigma}  f_{i\sigma}^{\dagger}
f^{\phantom{\dagger}}_{j\sigma}  +  
U \sum_i n_{i\uparrow}^{\phantom{\dagger}} n_{i\downarrow}^{\phantom{\dagger}} \,,
\end{equation}
that describes interacting fermions hopping on a lattice between nearest
neighbor sites with amplitude $-t$. The screened Coulomb interaction is
assumed local, and its strength on each site $i$ is given by $U$. It was later
on extended by Ole\'s to multiband systems to better embrace the diversity of
transition metal oxides \cite{Ole83}. A fundamental consequence of strong
correlations was already recognized by Hubbard, who showed that they split the
non-interacting tight-binding band  and give rise to additional features in
the excitation spectra including the upper Hubbard band (UHB)
\cite{Hub63,Hub64}. In fact, more recent investigations of the one-band
Hubbard Model within dynamical mean-field theory revealed that its
one-particle excitation spectra generically consist of lower and upper Hubbard
bands, together with a quasi-particle peak \cite{Geo96,Bul01}. These genuine
interaction-driven features are reflected in two-particle excitation
spectra. For instance, as shown in Fig.~\ref{fig:fs} the charge excitation
spectra consist of a continuum, a zero-sound collective mode, and a
high-frequency collective mode originating from the upper Hubbard band
\cite{Dao17}. The latter escapes a description within perturbation theory. 

The Kotliar and Ruckenstein slave-boson representation of the Hubbard model
\cite{Kot86} is a convenient tool to obtain these spectra at one-loop order in
the paramagnetic phase \cite{Dao17}. Such a symmetry-breaking free calculation
may be performed in the thermodynamical limit. It thus allows to resolve the
full momentum dependence of the spectra. Since incommensurate magnetic
instabilities are strongly suppressed with increasing temperature, neglecting
magnetic instabilities does not severely constrain the parameter range where
the calculation may be meaningfully performed. For instance, on the square
lattice, they essentially disappear for $T\approx t/6$ \cite{Dao17}. The
Gutzwiller approximation, which entails the interaction driven Brinkman-Rice
metal-to-insulator transition~\cite{Bri70}, is reproduced on the saddle-point
level by the Kotliar and Ruckenstein slave-boson
representations~\cite{Kot86,FW}. 

The zero-temperature Brinkman-Rice metal-to-insulator transition is second 
order~\cite{Bri70}. 
It turns first order at finite temperature as the latter may destabilize a poorly 
coherent Fermi Liquid \cite{Doll3, FK}. It is located in a coexistence region between 
a metallic and an insulating phase. Furthermore this first order transition extends 
to finite doping, where it is replaced by a transition from a good metal (with large 
quasi-particle residue) to a poor metal (with small quasi-particle residue). It 
extends up to a critical endpoint that depends on temperature \cite{Dao17, Mez17}. 
In contrast, a zero-temperature first order transition has been found in the two-band 
model in the vicinity of half-filling \cite{Fre02}.
The role of the lattice geometry on the metal-to-insulator transition 
was also investigated~\cite{Kot00}. In particular, a very good agreement 
on the location of the metal-to-insulator transition has been found with 
Quantum Monte Carlo simulations on the honeycomb lattice~\cite{Doll3}. 
Further comparisons of groundstate energies 
to existing numerical solutions have also been carried out for the square 
lattice. Regarding groundstate energies, for instance for $U=4t$, it could 
be shown that the slave-boson result is larger than its counterpart by less 
than 3\%~\cite{Fre91}. For larger values of $U$, the slave-boson groundstate 
energy exceeds the exact diagonalization data by less than 
4\% (7\%) for $U=8t$ ($20t$) and doping larger than 
15\%. The discrepancy increases when the doping is lowered~\cite{Fre92}. 
The saddle-point approximation is exact in the large degeneracy limit, 
and the Gaussian fluctuations provide $1/N$ corrections~\cite{FW}. 
Moreover it obeys a variational principle in the limit of large spatial 
dimensions where the Gutzwiller approximation becomes exact for the 
Gutzwiller wave function~\cite{Met89,Met88,Met89b}. 
Let us finally emphasize that a quantitative agreement was established 
between the charge structure factors calculated from Gaussian fluctuations 
within the slave-boson approach and quantum Monte Carlo 
simulations~\cite{Zim97}.

Numerous valuable results have been obtained with Kotliar and 
Ruckenstein~\cite{Kot86} and related slave-boson representations~\cite{Li89,FW}. Special
attention has been paid to anti-ferromagnetic~\cite{Lil90}, spiral~\cite{Fre91,Igo13,Fre92,Doll2}, and
striped~\cite{SeiSi,Fle01,Sei02,Sei03,Sei05,Rac06,Rac07,RaEPL} phases. In addition, 
the competition between the latter two has been addressed as
well~\cite{RaEPL}. It has also been obtained that the spiral order
continuously transforms to the ferromagnetic order in the large $U$ regime ($U
\gtrsim 60t$)~\cite{Doll2} so that its experimental realization is unlikely. 
Furthermore, in the two-band model on the square lattice, ferromagnetism was predicted 
as a possible groundstate in the doped Mott insulating regime only~\cite{Fre02}. 
However, the ferromagnetic instability line could be brought down to the intermediate
coupling regime when taking into account a ferromagnetic exchange 
coupling~\cite{lhoutellier15}. A sufficiently large next-nearest-neighbor hopping 
amplitude~\cite{FW98}, as well as going to the fcc lattice~\cite{Igo15}, results in a 
similar effect. In addition, this formalism extended to the Hubbard model with inter-site 
Coulomb interaction has been applied to address the strongly inhomogeneous 
polaronic states observed in correlated heterostructures~\cite{Pav06}. 
Most recently the possibility to enhance the capacitance by strong 
correlation effects in the metallic plates of a capacitor has been investigated 
within this approach~\cite{Ste17}.

Yet it needs to be verified that the approximate analytical expression used to
calculate the charge excitation spectra complies with the 
symmetries of the model. Clearly, translational invariance and spin-rotational
invariance are satisfied in a paramagnetic phase. However the less obvious
particle-hole symmetry remains to be established. 
Furthermore, there is a certain degree of arbitrariness inherent to
this representation as to how to perform the one-loop calculation: While the
internal gauge symmetry group of the representation allows to simplify the
problem, as the phase of three of the four slave-boson fields may be gauged
away by promoting the Lagrange multipliers to time-dependent fields, there is
no prescription to determine which one of them must remain a complex field. In
this paper, we not only show that the charge excitation spectra computed in
\cite{Dao17} are indeed particle-hole symmetric, but also that they do not
depend on whether the selected complex slave-boson field describes doubly
occupied sites or empty sites.

\section{Model and method} \label{sec:method}
In the spin-rotation invariant (SRI) Kotliar and Ruckenstein slave-boson 
representation~\cite{Kot86,fresard12} the Hubbard Hamiltonian is expressed with 
pseudo-fermion operators $f_{i\sigma}$ and auxiliary boson operators 
$e_i$, $p_{i\mu}$, $d_i$ (for atomic states with respectively zero, single and 
double occupancy) as
\begin{equation} \label{eq:model}
 H = -t \sum_{\langle i,j\rangle} \sum_{\sigma, \sigma', \sigma''} z_{i\sigma'' \sigma}^{\dagger} 
 f_{i\sigma}^{\dagger} f_{j\sigma'} z_{j\sigma' \sigma''} + U \sum_i d_i^{\dagger} d_i.
\end{equation}
In this form the on-site Coulomb interaction has the advantage to
be bilinear in bosonic operators. The canonical operators $p_{i\mu}$ 
build a $2\times2$ matrix in spin space in order to preserve spin rotation 
symmetry~\cite{Li89,FW}. It is expanded into the identity matrix 
$\underline{\tau}^0$ and the Pauli matrices as 
${\underline p}_i = \frac{1}{2} \sum_{\mu=0}^3 p_{i\mu} {\underline \tau}^{\mu}$. 
The operator $\underline{z}_i$ takes into account the occupancy change that occurs during
a hopping process. In the spin space it is also a matrix defined as
\begin{equation}
 {\underline z}_i = e_i^{\dagger} {\underline L}_i M_i {\underline R}_i \, 
 {\underline p}_i + {\underline {\tilde{p}}}_i^{\dagger} {\underline R}_i M_i  
 {\underline L}_i \, d_i
\end{equation}
with 
\begin{eqnarray}
& M_i & = \Big[ 1 + e_i^{\dagger} e_i + \sum_{\mu=0}^3 p_{i\mu}^{\dagger} p_{i\mu} 
+  d_i^{\dagger} d_i \Big]^{1/2}, \nonumber \\
&{\underline L}_i & = \Big[ (1 -d_i^{\dagger} d_i) {\underline \tau}^0 
- 2 {\underline p}_i^{\dagger} {\underline p}_i \Big]^{-1/2}, \nonumber \\  
& {\underline R}_i & = \Big[ (1 - e_i^{\dagger} e_i) {\underline \tau}^0 
- 2 {\tilde{\underline p}}_i^{\dagger} {\tilde{\underline p}}_i \Big]^{-1/2}          
\end{eqnarray}
where  $\tilde{\underline p}_i = \frac{1}{2} ( p_{i0} {\underline \tau}^0 
- {\bf p}_i \cdot \boldsymbol{\underline \tau} )$.

The subspace of physical states in the augmented Fock space generated by the 
auxiliary fermion and boson operators is the intersection of the kernels of 
operators 
\begin{eqnarray}\label{eq:const}
 & {\cali A}_i &= e_i^{\dagger} e_i + \sum_{\mu=0}^3 p_{i\mu}^{\dagger} p_{i\mu} 
 +  d_i^{\dagger} d_i - 1,  \nonumber\\
 & {\cali B}_{i0} & = \sum_{\mu=0}^3 p_{i\mu}^{\dagger} p_{i\mu} + 2 d_i^{\dagger} d_i 
 - \sum_{\sigma} f_{i \sigma}^{\dagger} f_{i \sigma}, \\
 & \pmb{\cali B}_i & =  p_{i0}^{\dagger} {\bf p}_i + {\bf p}_i^{\,\dagger} p_{i0} 
 - \ii {\bf p}_i^{\,\dagger} \times {\bf p}_i  - \sum_{\sigma, \sigma'} 
 \boldsymbol{ \tau}_{\sigma \sigma'} f_{i\sigma'}^{\dagger} f_{i\sigma}, \nonumber
\end{eqnarray}
{\it i.e.} in this subspace ${\cali A}_i = 0$ that is the constraint of one 
atomic state per site, and ${\cali B}_{i\mu} = 0$ which equates the 
number of fermions to the number of $p$ and $d$ bosons.

Functional integration is used to calculate the partition function~\cite{li91,Zim97} 
with the effective Lagrangian 
${\cal L} = {\cal L}^{\rm B} + {\cal L}^{\rm F}$. Here the purely 
bosonic part is
\begin{eqnarray}
 {\cal L}^{\rm B} = \sum_i  \bigg[  e_i^{\dagger} \partial_{\tau} e_i
 & + & \sum_{\mu=0}^3 p_{i\mu}^{\dagger} \partial_{\tau} p_{i\mu}  
 + d_i^{\dagger} (\partial_{\tau} + U) d_i  \nonumber \\
 & + & \alpha_i {\cali A}_i + \sum_{\mu=0}^3 \beta_{i\mu}  {\cali B}_{i\mu}^{\rm B} \bigg]
\end{eqnarray}
with ${\cali B}_{i\mu}^{\rm B}$ being the bosonic part of the operator 
${\cali B}_{i\mu}$. After the integration of the fermion fields the mixed fermion-boson 
part may be written as
\begin{eqnarray}
{\cal L}^{\rm F} = & - {\rm tr} \Big\{ \ln \Big[ (\partial_{\tau} - \mu +  \beta_{i0}) 
\delta_{\sigma\sigma'} \delta_{ij} + \boldsymbol{\beta}_i \cdot 
\boldsymbol{\tau}_{\sigma\sigma'} \delta_{ij} \nonumber \\
 & + t_{ij} \sum_{\sigma_1} z_{j\sigma\sigma_1}^{\dagger} z_{i\sigma_1\sigma'} \Big] \Big\},
\end{eqnarray}
with $\mu$ the chemical potential. 
 The constraints that define the physical states are enforced 
with Lagrange multipliers $\alpha_i$ and $\beta_{i\mu}$. The problem may be simplified 
thanks to the internal gauge symmetry group of the representation. Indeed 
by promoting the Lagrange multipliers to time-dependent fields~\cite{FW}, the 
phases of $e$ and $p_{\mu}$ can be gauged away. This leaves us with radial 
slave-boson fields~\cite{Fre01}. Their saddle-point values may be viewed as 
an approximation to their exact expectation values that are 
generically non-vanishing~\cite{Kop07}. In fact, disproving earlier 
claims~\cite{Ras88,Lav90,Li94}, the slave-boson field 
corresponding to double occupancy $d_i = d'_i + {\rm i} d''_i$ however 
has to remain complex as emphasized by several authors~\cite{Jol91,Kot92,FW}. 
Furthermore the dynamics of the, now, real $e_i$ and $p_{i\mu}$ fields drops out of 
${\cal L}^{\rm B}$ due to the periodic boundary conditions on boson fields.

Within the approximation of Gaussian fluctuations, the action is 
expanded to second order in field fluctuations 
\begin{eqnarray}
&\psi(k) = \big(\delta e(k),\delta d'(k),\delta d''(k),\delta p_{0}(k),
\delta \beta_{0}(k), \delta \alpha(k), \nonumber \\ 
& \delta p_{1}(k),\delta \beta_{1}(k),\delta p_{2}(k),\delta \beta_{2}(k),
\delta p_{3}(k), \delta \beta_{3}(k) \big)
\end{eqnarray}
around the paramagnetic saddle-point solution 
\begin{equation}
\psi_{\rm MF} = (e, d, 0, p_0,\beta_0,\alpha, 0,0,0,0,0,0)
\end{equation}
as
\begin{equation}
 \int d\tau {\cal L}(\tau) = {\cal S}_{\rm MF} + \sum_{k,\mu,\nu} \psi_{\mu}(-k) 
 S_{\mu\nu}(k) \psi_{\nu}(k)
\end{equation}
(the matrix $S$ is given in Appendix~A of \cite{Dao17}). We have introduced the 
notation $k=({\bf k},\nu_n)$, with the Matsubara frequencies $\nu_n=2\pi nT$,
and $\sum_k = T \sum_{\nu_n} L^{-1} \sum_{{\bf k}}$ with $L$ the number of  
lattice sites. The correlation functions of boson fields are then 
Gaussian integrals which can be obtained from the inverse of the fluctuation 
matrix $S$ as 
$\langle \psi_{\mu}(-k) \psi_{\nu}(k) \rangle = \frac{1}{2} S^{-1}_{\mu\nu}(k)$. 
Using the density fluctuation 
$\delta {\cali N} = \delta(d^{\dagger} d - e^{\dagger} e) $, 
the charge susceptibility is
\begin{eqnarray}
 \chi_c(k) & = & \langle \dd {\cali N}(-k) \dd {\cali N}(k) \rangle \nonumber \\
 &  = & 2 e^2 S_{1,1}^{-1}(k) - 4 e d S_{1,2}^{-1}(k) + 2 d^2  S_{2,2}^{-1}(k) .
 \label{eq:chi_c-def}
\end{eqnarray}
Dynamical response functions are eventually evaluated within analytical 
continuation ${\rm i} \nu_n \rightarrow \omega + {\rm i} 0^+$.

\section{Symmetry of the saddle-point solution} 

The Hubbard model possesses the particle-hole symmetry on a bipartite 
lattice. We show below that the symmetry is preserved at the saddle-point 
level in the specific case of the square lattice.

At first we present the general results of the paramagnetic 
solution, which do not presume any property of the lattice. At the 
saddle-point level, the boson values can be expressed with 
the hole doping from half-filling $\delta$ and the variable $x =e +d$ as  
\begin{equation}\label{eq:boson}
 e = \frac{x^2 + \delta}{2x}, \quad d = \frac{x^ 2 - \delta}{2 x} ,
\quad p_0^2 = 1 - \frac{x^4 + \delta^ 2}{2 x^2}. 
\end{equation}
The bare quasiparticle dispersion $t_{\bf k}$ is renormalized as 
\begin{equation}
 E_{{\bf k}} = z_0^2 t_{{\bf k}} - \mu_{\rm eff}
\end{equation}
with the factor 
 \begin{equation}\label{eq:z0}
 z_0^2= \frac{2 p_0^2 (e+d)^2}{1-\delta^2},
 \end{equation}
and $\mu_{\rm eff}$ the effective chemical potential.

The paramagnetic solution for fixed values of doping $\delta$ and 
coupling $U$ is found by determining the chemical potential 
$\mu_{\rm eff}$ via the filling condition
\begin{equation}
  \frac{2}{L} \sum_{{\bf k}} n_F(E_{{\bf k}}) = 1 - \delta
\end{equation}
 and the solution $x$ of the saddle-point equation
\begin{equation}\label{eq:saddle-point}
\frac{(1-x^2) x^4}{x^4 - \delta^2}  =  \frac{U}{U_0}.
\end{equation}
Here the coupling scale
\begin{equation}
 U_0 = - 8 \varepsilon_0/(1-\delta^2)
\end{equation}
has been introduced in terms of the semi-renormalized kinetic energy 
 \begin{equation}
 \varepsilon_0 =  \frac{2}{L} \sum_{{\bf k}}  t_{{\bf k}} n_F(E_{{\bf k}})
 \end{equation}
 and the Fermi function $n_F(\epsilon) = 1/(\exp(\epsilon/T) + 1)$. 

 Solving the saddle-point equation (Eq.~\ref{eq:saddle-point}) at half-filling yields 
$z_0^2=1-(U/U_0)^2$, in which case the effective mass of the quasiparticles
diverges when $U$ reaches $U_0$. This is the Brinkman-Rice mechanism of the
metal-to-insulator transition, as it also arises in the Gutzwiller 
approximation~\cite{Bri70,Vol84,Vol87}. 
 
In order to establish the particle-hole symmetry of the saddle-point 
approximation, we show that the solution for the opposite doping
is obtained by swapping the values of the empty and double occupancy, $e$ 
and $d$, and reversing the sign of the effective chemical potential 
$\mu_{\rm eff}$, while keeping unchanged $p_0$. Hence the saddle-point
value $x$ and the renormalization factor $z_0$ are even functions of 
$\delta$. The boson expressions~(\ref{eq:boson}) comply with the 
transformation, so it remains to check that the latter (i) leaves invariant 
the saddle-point equation and (ii) yields the filling condition for the 
opposite doping.  This can be achieved using the parity of the quasiparticle 
density of state $N(-\epsilon) = N(\epsilon)$.   
Alternatively, the dispersion on
the square lattice 
\begin{equation}
t_{{\bf k}} = -2t(\cos k_x +\cos k_y)
\end{equation}
yields the property $t_{{\bf k}+{\bf Q}} = - t_{\bf k}$, with ${\bf Q} =
(\pi,\pi)$, that will be used in the next section.

For (i) one can remark 
that both sides of Eq.~(\ref{eq:saddle-point}) are even in $\delta$ since
\begin{eqnarray}
 \varepsilon_0(-\delta) & = &\frac{2}{L} \sum_{{\bf k}}  t_{{\bf k}} n_F(z_0^2 t_{\bf k} 
 + \mu_{\rm eff})  \\
 & = &\frac{2}{L} \sum_{{\bf k'}}  t_{{\bf k'}+{\bf Q}} n_F(z_0^2 t_{{\bf k'}+{\bf Q}}
 + \mu_{\rm eff}) \nonumber \\
 & = & \frac{2}{L} \sum_{{\bf k'}} t_{\bf k'} \big[n_F(z_0^2 t_{{\bf k'}} - \mu_{\rm eff}) - 1\big]
  = \varepsilon_0(\delta) \nonumber
\end{eqnarray}
(the last equality results from the vanishing of $\sum_{\bf k} t_{\bf k}$ over the Brillouin zone)
so the saddle-point equation is invariant.
 As for (ii), the density of the transformed solution
\begin{eqnarray}
 \frac{2}{L} \sum_{{\bf k}} n_F(z_0^2 t_{{\bf k}} + \mu_{\rm eff}) 
 & = & \frac{2}{L} \sum_{{\bf k'}} n_F(z_0^2 t_{{\bf k'}+{\bf Q}} + \mu_{\rm eff}) \nonumber \\
 & = & \frac{2}{L} \sum_{{\bf k'}} \big[ 1 - n_F(z_0^2 t_{{\bf k'}} - \mu_{\rm eff}) \big] \nonumber \\
 & = & 2 - (1 - \delta) = 1 + \delta
\end{eqnarray}
indeed corresponds to the opposite doping.

\section{Symmetry of the quasiparticle response functions}
On the square lattice the quasiparticle response function is transformed
under the reversal of the doping sign as
\begin{equation}
 \chi_m^-(k) = (-1)^m \chi_m^+(k). 
\end{equation}
Here we have introduced the notation for the generalized quasiparticle 
response functions at doping $\pm\delta$ 
\begin{equation}
 \chi_m^{\pm}(k) = \frac{2}{L} \sum_{{\bf q}}  (t_{{\bf q} + {\bf k}} + t_{\bf q} )^m 
 \frac{n_F(E_{{\bf q} + {\bf k}}^{\pm}) - n_F(E_{\bf q}^{\pm})}{\omega - 
 (E_{{\bf q} + {\bf k}}^{\pm} - E_{\bf q}^{\pm}) } 
\end{equation}
with $E_{\bf q}^{\pm} = z_0^2 t_{\bf q} \mp \mu_{\rm eff}$. 
The relation between the expressions at opposite dopings can be derived by 
summing instead on ${\bf p} = -{\bf q} - {\bf k} + {\bf Q}$. This yields
\begin{eqnarray}
 \chi_m^-(k) & = &\frac{2}{L} \sum_{{\bf p}}  (t_{{\bf -p} + {\bf Q} } 
 + t_{-{\bf p}-{\bf k}+ {\bf Q}} )^m \\
 & & \times \frac{n_F(E_{-{\bf p}+ {\bf Q}}^-) - n_F(E_{-{\bf p}-{\bf k}+ {\bf Q}}^-)}
 {\omega - (E_{-{\bf p}+ {\bf Q}}^- - E_{-{\bf p}-{\bf k}+ {\bf Q}}^-) } \nonumber \\
 & = & \frac{2}{L} \sum_{{\bf p}}  (- t_{{\bf p}} - t_{{\bf p}+{\bf k}} )^m 
 \frac{n_F(- E_{{\bf p}}^+) - n_F(-E_{{\bf p}+{\bf k}}^+)}
 {\omega - (- E_{{\bf p}}^+ + E_{{\bf p}+{\bf k}}^+) } \nonumber
\end{eqnarray}
wherein we have used the equalities $t_{-{\bf p} + {\bf Q}} = -t_{\bf p}$ 
and $E_{-{\bf p} + {\bf Q}}^{-} = -z_0^2 t_{\bf p} + \mu_{\rm eff} = - E_{\bf p}^{+}$.
Finally the sum can be written as
\begin{equation}
 \chi_m^-(k) = (-1)^m \frac{2}{L} \sum_{{\bf p}}  (t_{{\bf p}+{\bf k}} + t_{\bf p} )^m 
 \frac{n_F(E_{{\bf p}+{\bf k}}^+) - n_F(E_{{\bf p}}^+)}
 {\omega - (E_{{\bf p}+{\bf k}}^+ - E_{{\bf p}}^+) } 
\end{equation}
since $n_F(-\epsilon) = 1 - n_F(\epsilon)$.

{\it En passant} the above relation shows the particle-hole symmetry of the RPA charge 
response function
\begin{equation}
\chi_{\rm RPA}(k) = \frac{\chi_0^{(0)}(k)}{1 + \frac{U}{2} \chi_0^{(0)}(k)}
\end{equation}
where $\chi_0^{(0)}(k)$ is the charge response function of a Fermi gas, 
{\it i.e.} $\chi_0^{(0)}(k) = \chi_0(k)|_{z_0 = 1}$.

\section{Symmetry of the slave-boson charge response}

The charge dynamical response function has been given in Ref.~\cite{Dao17}. In
order to demonstrate its particle-hole symmetry, we cast it into an expression
which is explicitly invariant under the transformation $\{ \delta \mapsto
-\delta, e \mapsto d, d \mapsto e, \chi_m(k) \mapsto (-1)^m \chi_m(k) \}$
undergone when reversing the doping sign. After a lengthy but straightforward
expansion, it can be written as 
\begin{equation}\label{eq:chi_c}
 \chi_c(k) = \frac{A(k) + B(k)  \omega^2}{C(k) + D(k)  
 \omega^2}
\end{equation}
where
\begin{align}\label{eq:coeff}
& A(k) = \frac{ 2  p_0^2 \varepsilon_0 }{1 - \delta^2}  \bigg[ \Big( p_0^2(d^2 S_{11} + e^2 S_{22}) + 4 e^2 d^2 S_{44} \nonumber \\
& \qquad \qquad \quad + 2edp_0^2 S_{12} - 4edp_0(d S_{14} + e S_{24})  \Big) \chi_0(k) \nonumber \\
& \qquad \qquad \quad  + 2 \big( e \Delta_1 - d \Delta_2 \big)^2 \bigg], \nonumber \\
& B(k) = ed p_0^2 \chi_0(k), \nonumber \\
& C(k) = \frac{ 2  p_0^2 \varepsilon_0 }{1 - \delta^2}   \bigg[
\bigg( p_0^2 S_{11} S_{22} - (p_0 S_{12} - d S_{14} - e S_{24})^2  \nonumber \\
& \qquad \qquad \quad + 4 ed S_{14} S_{24} - 2 p_0( d S_{11} S_{24} + e S_{22} S_{14})  \nonumber \\
& \qquad \qquad \quad + (d^2 S_{11} + e^2 S_{22} - 2ed S_{12}) S_{44} \bigg) \frac{\chi_0(k)}{2} \nonumber \\
& \qquad \qquad \quad + S_{11} \Delta_{1}^2 + S_{22} \Delta_{2}^2 + S_{44} \Delta_{4}^2  + 2 S_{12} \Delta_{1} \Delta_{2}
 \nonumber \\
& \qquad \qquad \quad -  2 S_{14} \Delta_{1} \Delta_{4} - 2 S_{24} \Delta_{2} \Delta_{4} \bigg], \nonumber \\
& D(k) = \frac{ed}{(e+d)^2} \bigg[ \quad \bigg( p_0^2 (S_{11} + S_{22}) + (e-d)^2 S_{44} \nonumber \\
& \qquad \qquad \quad - 2 p_0^2 S_{12} + 2 (e-d) p_0 (S_{24} - S_{14}) \bigg) \frac{\chi_0(k)}{2} \nonumber \\
& \qquad \qquad \quad + \big( \Delta_1 + \Delta_2 \big)^2 \quad \bigg].
\end{align}
Here the elements of the fluctuation matrix are
\begin{align}
S_{11} & = -\varepsilon_0 \frac{z_0}{e} \frac{\partial z}{\partial e} + s_{11}(k), \nonumber \\
S_{22} & = -\varepsilon_0 \frac{z_0}{d} \frac{\partial z}{\partial d'} + s_{22}(k), \nonumber \\
S_{44} & = -\varepsilon_0 \frac{z_0}{p_0} \frac{\partial z}{\partial p_0} + s_{44}(k), \nonumber \\
S_{\mu \nu} & = s_{\mu\nu}(k) \;\;  \text{for $\mu,\nu=1,2,4$ with $\mu\neq \nu$} 
\nonumber \\
\end{align} 
where
\begin{equation}
 s_{\mu\nu}(k) = \varepsilon_0 z_0 \frac{\partial^2 z}{\partial \psi_{\mu} \partial \psi_{\nu}} 
 + \frac{\partial z}{\partial \psi_{\mu}} \frac{\partial z}{\partial \psi_{\nu}} 
 \left[ \varepsilon_{{\bf k}} - \frac{z_0^2}{2}  \chi_2(k) \right] 
\end{equation}
with \begin{equation}
 \varepsilon_{\bf k} = \frac{2}{L} \sum_{{\bf q}} t_{{\bf q}+{\bf k}} n_F(E_{\bf q}),
 \label{eq:epsk}
\end{equation} 
and
\begin{align}
& \Delta_{1} =  d p_0 + \left( p_0 \frac{\partial z}{\partial d'} - d \frac{\partial z}{\partial p_0} \right)
\frac{z_0}{2} \chi_1(k), \nonumber \\
 & \Delta_{2} =  e p_0 - \left( p_0 \frac{\partial z}{\partial e} - e \frac{\partial z}{\partial p_0} \right)
 \frac{z_0}{2} \chi_1(k),  \nonumber \\
 & \Delta_{4} = 2 e d + \left( e  \frac{\partial z}{\partial d'} - d \frac{\partial z}{\partial e} \right)
 \frac{z_0}{2} \chi_1(k).  
\end{align}

The expressions of the derivatives of $z$ may be gathered from Ref.~\cite{li91,Zim97}. 
The first-order derivatives are
\begin{align}
 \frac{\partial z}{\partial e}  & = \sqrt{2} \eta p_0 \left( 1 + \frac{2 x e}{1-\delta} \right),  \nonumber \\
 \frac{\partial z}{\partial d'} & = \sqrt{2} \eta p_0 \left( 1 + \frac{2 x d}{1+\delta} \right), \nonumber \\
 \frac{\partial z}{\partial p_0} & = \sqrt{2} \eta x \left( 1 + \frac{2 p_0^2}{1-\delta^2} \right),
\end{align}
with $\eta =1/\sqrt{1-\delta^2}$. The second-order derivatives are
\begin{align}
 \frac{\partial^2 z}{\partial e^2} & = \frac{2\sqrt{2}\eta p_0}{1-\delta} \left( x + 2e 
 + \frac{6 x e^2}{1-\delta} \right), \\
  \frac{\partial^2 z}{\partial d'^2} & = \frac{2\sqrt{2}\eta p_0}{1+\delta} \left( x + 2d 
 + \frac{6 x d^2}{1+\delta} \right), \nonumber \\
 \frac{\partial^2 z}{\partial p_0^2} & = 2\sqrt{2}\eta^3 x p_0 \Big(3 + (6\eta^2 - 2) p_0^2 \Big), \nonumber \\ 
 \frac{\partial^2 z}{\partial e \partial d'} & = 2\sqrt{2}\eta p_0 \left( \frac{e}{1-\delta} 
 + \frac{d}{1+\delta} + 2 \eta^2 x e d \right), \nonumber \\ 
 \frac{\partial^2 z}{\partial e \partial p_0} & = \sqrt{2}\eta \left( 1 + 2\eta^2 p_0^2 (1 + x e)
 + \frac{2 x e}{1-\delta} + \frac{6 x e p_0^2}{(1-\delta)^2} \right), \nonumber \\ 
 \frac{\partial^2 z}{\partial d' \partial p_0} & = \sqrt{2}\eta \left( 1 + 2\eta^2 p_0^2 (1 + x d)
 + \frac{2 x d}{1+\delta} + \frac{6 x d p_0^2}{(1+\delta)^2} \right). \nonumber
\end{align}

It should be emphasized that due to the symmetry of the saddle-point solution,
{\it i.e.}  
$e({-\delta}) = d({\delta})$ and $p_0(-\delta) = p_0(\delta)$, the values of the partial 
derivatives of $z$ by $e$ are interchanged with those by $d'$ when reversing
the sign of the doping, {\it e.g.} $(\partial^2 z/\partial e\partial p_0) (-\delta) = (\partial^2 z/\partial 
d'\partial p_0) (\delta)$. The values of the fluctuation matrix 
elements $S_{\mu\nu}$ with indices $1$ and $2$ are thus interchanged, {\it e.g.} 
$S_{11}(-\delta) = S_{22}(\delta)$ or $S_{14}(-\delta)= S_{24}(\delta)$. 
Furthermore one can check that $\Delta_4(-\delta) = \Delta_4(\delta)$ and 
$\Delta_{1}(-\delta) = \Delta_{2}(\delta)$. As a result, the coefficients~(\ref{eq:coeff}), 
and then the response function, are actually invariant. 

\begin{figure}[h]
  \includegraphics[clip=true, width=0.48\textwidth]{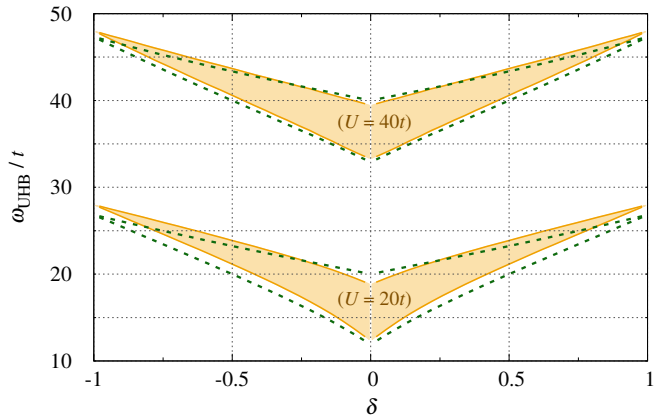}
  \caption{(Color online) Dispersion of the UHB mode following from
    Eq.~(\ref{eq:chi_c}) (shaded area) as a 
    function of the doping for $U=20t$ and $40t$. The dashed boundaries are
    obtained from the strong-coupling approximation Eq.~(\ref{eq:wuhb2}).     
  Parameter: $T=t/100$.}
  \label{fig:wid}
\end{figure}

As an example we plotted the numerical evaluation of Eq.~(\ref{eq:chi_c}) in
Fig.~\ref{fig:fs}, where the characteristic features of the inelastic charge
response function are clearly visible. While neither the continuum nor the
zero-sound collective mode above the upper edge of the continuum could be
brought into a simple analytical form, an approximate expression of the
UHB pole for all momenta applicable in the strong coupling regime could be
derived \cite{Dao17}. It reads
\begin{equation}
 \omega_{\rm UHB}({\bf k}) \approx U \sqrt{1 - \frac{U_0}{2U} \left( 1-3|\delta| 
 + (1-|\delta|) \frac{\varepsilon_{\bf k}}{\varepsilon_0}\right)}\,. 
  \label{eq:wuhb2}
\end{equation}
In Fig.~\ref{fig:wid} we compare the evaluation of Eq.~(\ref{eq:chi_c}) with
the approximation Eq.~(\ref{eq:wuhb2}) for $U=20t$ and $40t$. The smallest
energy of the UHB mode peak is obtained for ${\bf k} = (0,0)$, and the highest
one for ${\bf k} = (\pi,\pi)$. The dispersion of this mode is approximately
given by $-2 (1-|\delta|) \varepsilon_{\bf k}$, and is accordingly largest at
half-filling and vanishingly small in the limit of empty/full
system. Furthermore, the difference between the approximation
Eq.~(\ref{eq:wuhb2}) and Eq.~(\ref{eq:chi_c}) tends to vanish already for
$U=40t$.

Let us finally mention an alternative derivation of Eq.~(\ref{eq:chi_c}). We
recall that there is some arbitrariness to its derivation as we chose here to
gauge away the phase of the $e$-boson (on top of the one of the
$p_{\mu}$-bosons). Alternatively one may chose to gauge away the phase of the
$d$-boson while keeping the $e$-boson as a complex field. This obviously leads
to a $S$-matrix that differs from the one derived in \cite{Dao17} and used
here. Yet, tedious work shows that $\chi_c(k)$ obtained this way is
nevertheless given by Eq.~(\ref{eq:chi_c}), thereby overcoming the above
mentioned arbitrariness and putting Eq.~(\ref{eq:chi_c}) on a firmer ground. 

\section{Conclusion}

Summarizing, the particle-hole symmetry of the charge response function
obtained for the Hubbard model using various approximations has been
considered. It has first been established that this symmetry is obeyed in the
random phase approximation for a bipartite lattice such as the square lattice
that we explicitly addressed. We then considered the expression of $\chi_c$
resulting from the SRI Kotliar and Ruckenstein slave boson representation
calculated to one-loop order again on the square lattice. In this case we
succeeded to cast its rather involved expression into a form that is manifestly
particle-hole symmetric. The latter also applies to the Kotliar and
Ruckenstein representation. Our arguments can easily be extended to other
bipartite lattices with one atom in the unit cell, such as the simple cubic one.

\section*{Acknowledgments}

The authors acknowledge the financial 
support of the French Agence Nationale de la Recherche (ANR), through
the program Investissements d'Avenir (ANR-10-LABX-09-01), LabEx EMC3, the
R\'egion Basse-Normandie, the R\'egion Normandie, and the Minist\`ere de la
Recherche.

\end{document}